\documentclass[11pt]{article}
\usepackage{moriond}
\usepackage{amssymb}
\usepackage{amsmath}

\def\qp{ {\bf q}_T }

\def\pp{ {\bf p}_T }
\def\kp{ {\bf k}_T }

\def\be{\begin{equation}}
\def\ee{\end{equation}}
\def\bea{\begin{eqnarray}}
\def\eea{\end{eqnarray}}

\newcommand{\Photo}{}

\begin{document}
\vspace*{4cm}
\title{Long-range angular correlations by strong color fields in hadronic collisions}

\author{Kevin Dusling}

\address{Physics Department, North Carolina State University, Raleigh, NC 27695, USA}

\maketitle\abstracts{
I present an overview of the {\em ridge} phenomenon in proton-proton and
proton-lead collisions.  This novel collimation between rapidity separated hadron
pairs is a consequence of non-linear gluon dynamics within the
small$-x$ wave-function of the colliding hadrons.
}

\begin{figure}[h!]
\centerline{\includegraphics[width=0.9\linewidth]{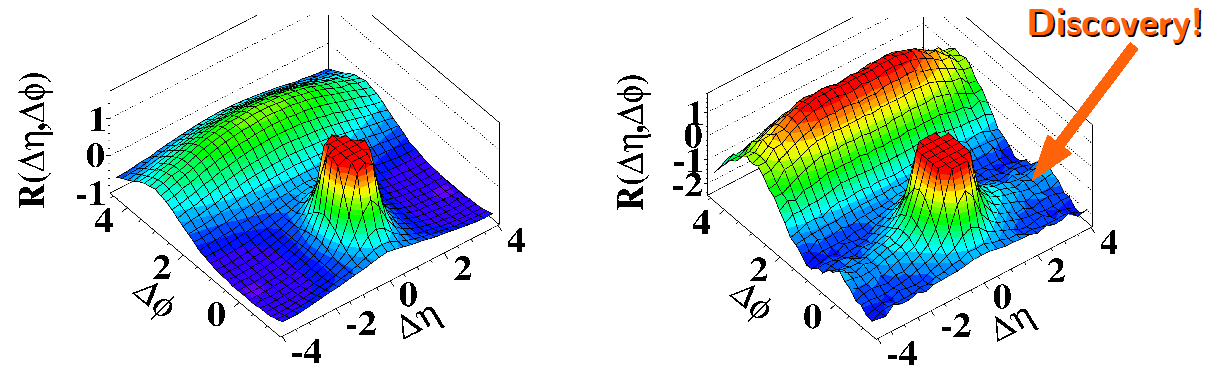}}
\caption{Two particle correlation function from minimum-bias (left) and
high-multiplicity (right) events in proton-proton collisions at $\sqrt{s}=2.76$~TeV.
Data from the CMS collaboration~\protect\cite{Khachatryan:2010gv}.}
\label{fig:ridgeA}
\end{figure}

Multi-particle correlations in high-energy hadronic collisions can serve as
sensitive tests of QCD dynanmics.  In the fall of 2010 the CMS collaboration
made a striking discovery~\cite{Khachatryan:2010gv}.  In high-multiplicity proton-proton collisions a
correlation was uncovered between charged-particle pairs having a large
pseudo-rapidity separation ($\Delta\eta \gg 1$)
and a narrow angular separation ($\Delta\phi\lesssim 1$) in azimuth.  
Figure~\ref{fig:ridgeA} displays the original CMS result.  As the structure
of the two-particle correlation resembles a mountain range the new discovery was dubbed 
``The Ridge" when it was first observed in heavy-ion collisions.  
It should be stressed that the heavy-ion ridge has a unique physical interpretation from the ridge in p+p and p+Pb collisions.  While the former is attributed primarily to final-state rescattering and hydrodynamic flow the latter, which will be the focus of this proceeding, is a consequence of gluon saturation~\cite{Dusling:2012iga}.

\begin{figure}
\centerline{\includegraphics[width=0.8\linewidth]{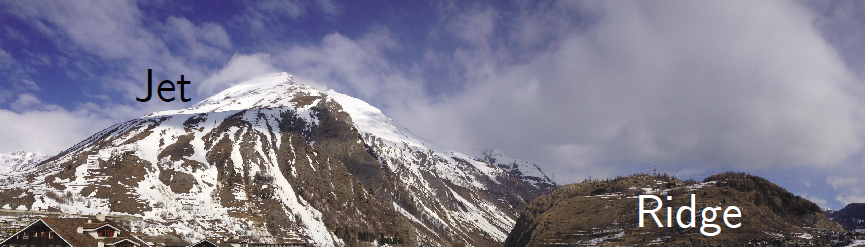}}
\caption{The ridge in La Thuile on March 14$^{\rm th}$ 2013.}
\label{fig:lathuile}
\end{figure}

\begin{figure}[h!]
\centerline{\includegraphics[width=0.9\linewidth]{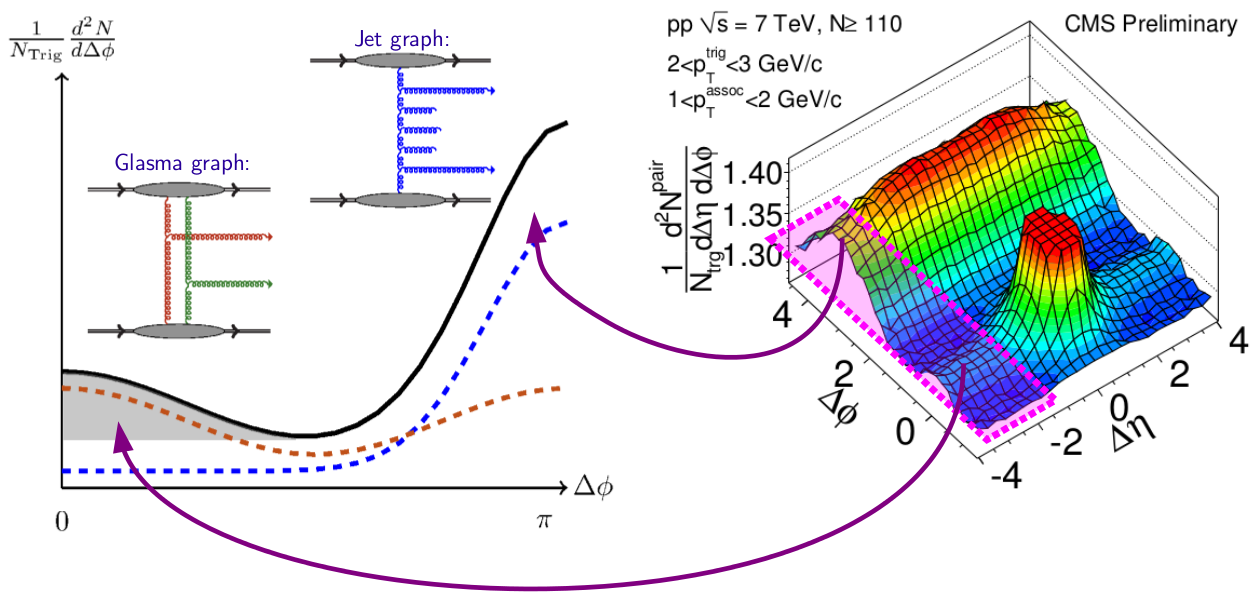}}
\caption{Anatomy of a proton-proton collision.  The away-side peak, associated
with mini-jet production arises from the jet-graph (two gluons produced from a single ladder) as shown
in the right diagram along with its schematic contribution to the
per-trigger-yield plotted in blue.  The Glasma graph contribution (left diagram) is shown
schematically by the orange curve.  The shaded gray region (extracted
experimentally by the ZYAM procedure) is referred to as the associated yield.}
\label{fig:ridgeB}
\end{figure}

The ridge in proton-proton and proton-lead collisions is a
natural consequence of gluon-saturation and non-linear parton dynamics in the nuclear
wave-function {\em before} the collision occurs.  The full structure of the
two-particle correlation is explained by two competing QCD diagrams; as
long as they are both computed within the CGC-EFT~\cite{Gelis:2010nm}.  The first diagram
corresponds to the production of two gluons from a single {\em ladder} as shown
in the right-most Feynman diagram appearing in Fig.~\ref{fig:ridgeB}.  This production
mechanism creates particles primarily back-to-back ({\em i.e.} on the
away-side for $\Delta\eta\gtrsim 1$) and does not generate any near-side collimation.  A second class of
diagrams, herein called ``Glasma Graphs", where two gluons are produced from
different ladders, is responsible for the near-side collimation identified with the
ridge.  The two-particle correlation generated by the Glasma graphs is symmetric with respect to $\Delta\phi=\pi/2$ and therefore produces an equal away-side collimation as well;
however, this is hidden underneath the away-side jet and was only recently observed after a careful subtraction of peripheral events.

Saturation dynamics comes into play when determining the relative contribution
of these two diagrams towards the two-particle correlation.  A power counting
exercise demonstrates that the Glamsa graph is enhanced by a factor of
$\alpha_S^{-8}$ in the presence of saturation, which we associate with
central / high-multiplicity collisions.  This should be compared to the 
$\alpha_S^{-4}$ enhancement of the jet-graph in going from min-bias to central
collisions.

The near-side collimation produced by the Glasma graph can be understood by looking at
the schematic form of the double-inclusive cross section
\begin{equation}
d^2\sigma \propto \int_{\kp}\Phi_A^2(x_1,\kp)\Phi_B(x_1,\pp-\kp)\Phi_B(x_2,\qp-\kp)+\cdots
\label{eq:1}
\end{equation}

The full expression used in computations along with all relevant details is presented in~\cite{Dusling:2012cg}.
The function $\Phi(x,\kp)$ appearing in these expressions is the unintegrated gluon
distribution and is obtained by solving the rcBK equation with MV model
initial conditions at large $x$.   The Cauchy-Schwarz inequality dictates that the correlation
strength at $\Delta\phi=0$ must be greater than or equal to that at
$\Delta\phi=\pi/2$; 
\begin{equation}
\int d^2 k_\perp \Phi_A^2 (\kp)\; \Phi_B^2 (\vert\pp -\kp\vert)  \geq \int d^2 k_\perp \Phi_A^2 (\kp)\; \Phi_B (\vert\pp -\kp\vert)\; \Phi_B (\vert\qp - \kp\vert) \;.
\end{equation}

\begin{figure}
\centerline{
\includegraphics[width=0.45\linewidth]{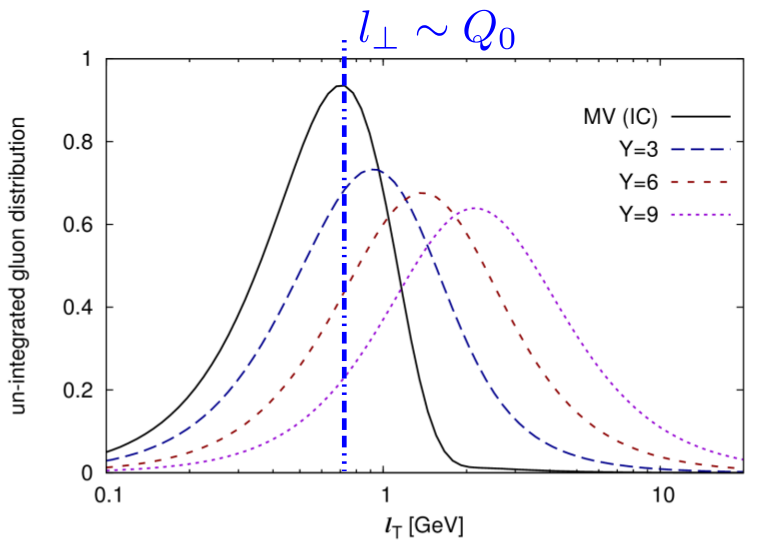}
}
\caption{Unintegrated gluon distribution function evolved from the M.V. model initial
condition\newline to larger rapidities via the rcBK evolution equation.}
\label{fig:ugd}
\end{figure}

Furthermore the equality holds if and only if
\begin{equation}
\Phi(\vert\pp - \kp\vert) \propto \Phi(\vert\qp - \kp\vert) \;,
\end{equation}
and therefore, as long as the gluon distribution is a non-trivial function of
momentum, there must be a 
collimation.  The strength depends on the detailed structure of the gluon
distribution function shown in fig.~\ref{fig:ugd}.

The rapid rise of the associated yield (see Fig.~\ref{fig:ridgeB} for the
definition) with centrality for asymmetric collisions can also be understood
from equation~\ref{eq:1} as well.  If we take the unintegrated gluon distribution function
as narrowly peaked around the saturation scale $Q_A$ and $Q_B$ of the projectile and
target respectively, the associated yield increases quadratically with the larger
of the two saturation scales~\cite{Dusling:2012wy},
\begin{equation}
\textrm{CY}\propto 
\frac{\Phi_B (Q_B)}{\Phi_B\left(\sqrt{2p_T^2+2Q_A^2-Q_B^2}\right) }
\xrightarrow[Q_B \gg Q_A]{} 1+\frac{Q_B^2}{Q_A^2}\;.
\end{equation}

\begin{figure}
\centerline{
\includegraphics[width=0.45\linewidth]{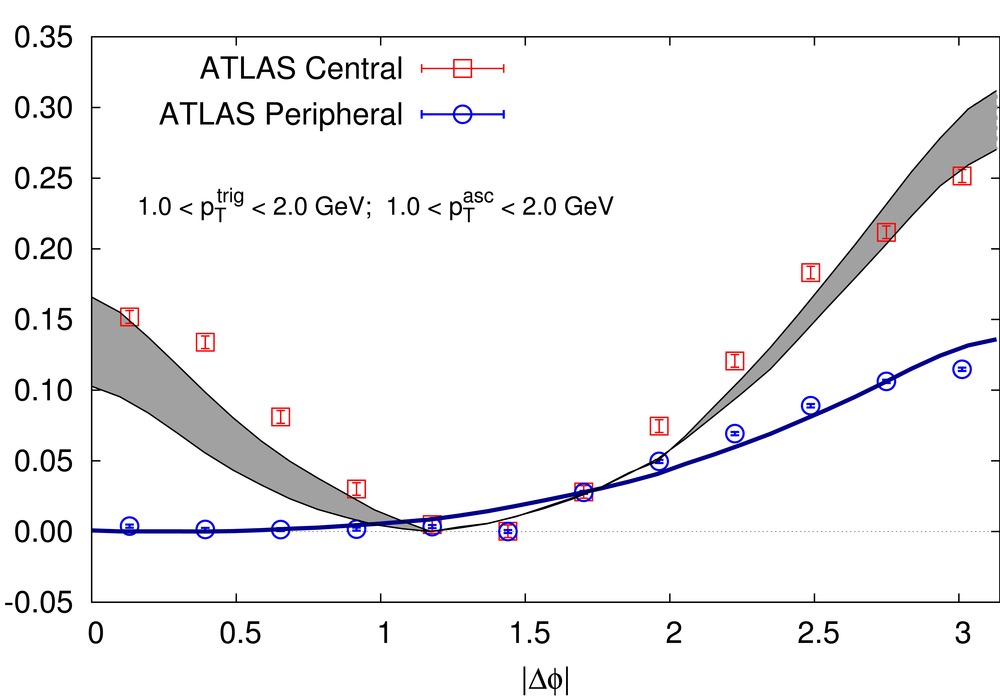}
\hspace{1cm}
\includegraphics[width=0.45\linewidth]{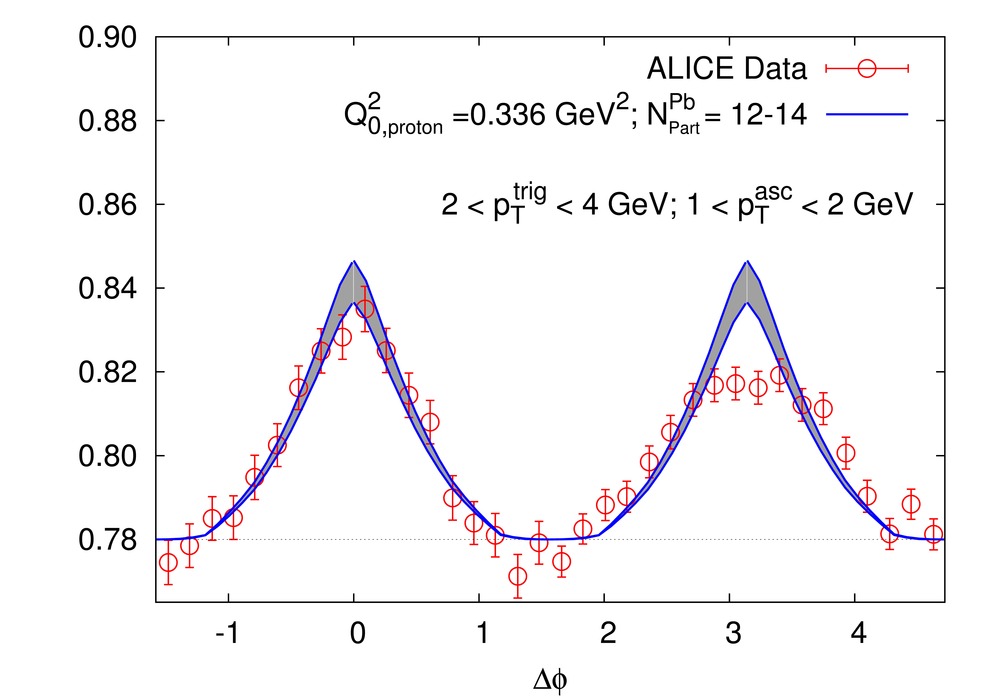}
}
\caption{Comparison of calculations within the CGC-EFT framework with recent
experiments.  The left figure shows the ATLAS~\protect\cite{Aad:2012gla}
per-trigger-yield as a function of azimuthal separation ($\Delta\phi$) for
peripheral events as blue circles and central events as red squares.  The solid
curves show the theory calculation for these two event classes.  For peripheral
events only the jet diagram is visible in both the data and theory calculation.
The right figure shows data from the ALICE collaboration~\protect\cite{Abelev:2012cya} where the difference between central and peripheral events allowed for the discovery of the away-side ridge.  The gray band shows shows the contribution from the Glasma graphs.}
\label{fig:atlas}
\end{figure}

A recent work~\cite{Dusling:2013oia} has compared predictions made within the
CGC-EFT to the CMS~\cite{CMS:2012qk},
ALICE~\cite{Abelev:2012cya} and ATLAS~\cite{Aad:2012gla} experimental data.  Two examples of the quality of the theory to data
comparison is shown in figure~\ref{fig:atlas}.  While the CGC-EFT provides a unified description of both p+p and p+Pb collisions at a quantitative level one may still question the role played by final state effects.  While this can only be addressed by detailed modeling there is strong evidence that final state interactions are small and not responsible for the ridge.  

A crucial piece of evidence stems from the away-side jet, which remains unmodified from peripheral to central collisions; a fact confirmed a posteriori by the successful subtraction of the peripheral jet from central data by the LHC experimentalists.  If final-state rescattering was strong enough to collimate the medium's particles to create the ridge, rescattering should similarly modify the jet.  This is indeed observed in heavy-ion collisions but it is not the case in p+p and p+Pb collisions.

A natural question is whether there is a unique final state interpretation of {\em both} the proton-proton and proton-lead ridge?
The differences in systematics between these two systems is naturally understood within the CGC framework.  Whether there is a consistent final-state model is far from clear.  It seems extremely difficult for a final-state interpretation to account for an associated yield four times larger in p+Pb than in p+p at the {\em same} multiplicity, considering our expectation that the overlap areas in the two systems are approximately the same.

There is compelling evidence that the ridge in proton-proton and proton-lead collisions is a consequence of gluon saturation and non-linear gluon dynamics.  This remarkable discovery of a novel collimation between
two particles flying in opposite directions in ultra-rare high multiplicity events is probing rare quantum fluctuations within the proton and nucleus at the smallest length scales experimentally possible.

\section*{Acknowledgments}

Supported by the US Department of Energy under Contract No. DE-FG02-03ER41260.

\end{document}